\documentclass[a4paper,english,conference]{IEEEtran}
\usepackage{latexsym}
\usepackage{float}
\usepackage{amsfonts}
\usepackage{amsbsy}
\usepackage{amssymb}
\usepackage{times}
\usepackage{graphicx}
\usepackage{enumerate}
\usepackage[usenames]{color}
\usepackage[dvips]{pstcol}
\usepackage{epstopdf}
\usepackage{cite}
\usepackage{amssymb}
\usepackage{amsfonts}
\usepackage{graphicx}
\usepackage{epsfig}
\usepackage{psfrag}
\usepackage{xcolor}
\usepackage{amsfonts, bm}
\usepackage{epstopdf}
\usepackage{cite}
\usepackage{color}
\usepackage{xcolor}
\usepackage{subfig}
\usepackage{verbatim}
\usepackage{multirow}
\usepackage{array}
\usepackage{booktabs}
\usepackage{amsthm}
\usepackage{makecell}

\usepackage[linesnumbered, ruled]{algorithm2e}
\usepackage{algpseudocode}
\usepackage{amsmath}

\IEEEoverridecommandlockouts

\begin{document}
	
	\title{Adaptive CSI Feedback for Deep Learning-Enabled Image Transmission \thanks{This work was supported by the National Natural Science Foundation of China under Grants 61971376, U22A2004, and 61831004, the Fundamental Research Funds for the Central Universities226-2022-00195, and the Defense Industrial Technology Development Program under Grant JCKY2020210B021.}
	\thanks{This paper has been accepted at IEEE ICC, Rome, Italy, May
		2023.}}
	\author{\IEEEauthorblockN{ Guangyi Zhang, Qiyu Hu, Yunlong Cai, and Guanding Yu   }
		\IEEEauthorblockA{College of Information Science and Electronic Engineering, Zhejiang University, Hangzhou, China\\
		Zhejiang Provincial Key Laboratory of Information Processing, Communication and Networking (IPCAN), \\ Hangzhou 310007, China }
		\IEEEauthorblockA{  E-mail: \{zhangguangyi, qiyhu, ylcai, yuguanding\}@zju.edu.cn} }
	
	\maketitle
	\vspace{-3.3em}
	\begin{abstract}
	Recently, deep learning-enabled joint-source channel coding (JSCC) has received increasing attention due to its great success in image transmission. However, most existing JSCC studies only focus on single-input single-output (SISO) channels. In this paper, we first propose a JSCC system for wireless image transmission over multiple-input multiple-output (MIMO) channels. As the complexity of an image determines its reconstruction difficulty, the JSCC achieves quite different reconstruction performances on different images. Moreover, we observe that the images with higher reconstruction qualities are generally more robust to the noise, and can be allocated with less communication resources than the images with lower reconstruction qualities. Based on this observation, we propose an adaptive channel state information (CSI) feedback scheme for precoding, which improves the effectiveness by adjusting the feedback overhead. In particular, we develop a performance evaluator to predict the reconstruction quality of each image, so that the proposed scheme can adaptively decrease the CSI feedback overhead for the transmitted images with high predicted reconstruction qualities in the JSCC system. We perform experiments to demonstrate that the proposed scheme can significantly improve the image transmission performance with much-reduced feedback overhead. 
	
	\end{abstract}
	

	\IEEEpeerreviewmaketitle
	
	\section{Introduction}
	
	 According to Shannon's separation theorem \cite{Shannon, 2006Elements}, modern data transmission is divided into a two-step separated encoding process,  namely source coding and channel coding. It has then been proven that the separated source and channel coding approaches its optimum theoretically when the blocklength goes to infinity. However, in the finite blocklength scenarios, the joint source-channel coding (JSCC) has been proven to achieve better performance than the separated scheme \cite{2005Joint}. Recently, inspired by the success of deep learning (DL), the autoencoder architecture parameterized by the neural networks (NNs) is used to implement the JSCC system, which outperforms the separated scheme\cite{JSCC,JSCCf,DeepSC,UDeepSC, OFDMJSCC, vqvae}. It employs deep neural networks (DNNs) to map the input source data directly to channel symbols in a joint manner, which is then decoded by another DNNs at the receiver. Specifically, in \cite{JSCC}, the authors firstly proposed the deep JSCC (D-JSCC) technique for wireless image transmission, where the image pixel values are mapped to the complex-valued channel symbols with a well-designed encoder. The authors in \cite{JSCCf} incorporated the channel output feedback into the D-JSCC system to improve the reconstruction performance. Moreover, a unified joint source-channel coding semantic communication system for multi-modal data transmission has been proposed in \cite{UDeepSC}.

	Multiple-input multiple-output (MIMO) has been widely deployed in practical communication systems for transmitting various sources, such as image, text, and video. It is regarded as a critical technology for current and future wireless systems, since it can provide high spectral efficiency and reduce the interference by fully utilizing the spatial resources \cite{CSISurvey}. However, these strengths are highly dependent on the available channel state information (CSI) at the base station (BS), and the user equipment is required to feed the CSI back to the BS through feedback links. Besides, the substantial antennas at the BS for massive MIMO lead to a huge dimension of the CSI matrix, which seriously increases the feedback overhead. Based on compressive sensing (CS), several algorithms have been developed to compress the CSI matrix to reduce the overhead \cite{AMP,CSCSI}. The authors in \cite{CSCSI} use the spatial correlations among nearby antennas to compress the CSI in the spatial-frequency domain. Moreover, there have been also many DL-based methods \cite{csi_cnn, Co_jiajia, CSIPaper1, CLNET} to compress the CSI matrix with the NNs, such as CsiNet \cite{CSIPaper1} and CLNet \cite{CLNET}. 
	These methods employ the autoencoder structure that comprises with encoder and decoder. Specifically, the encoder is deployed to sense and compress CSI into a low-dimensional codeword vector. Then, the decoder uses the received codeword vector to recover the original CSI matrix. 
	
	Existing D-JSCC methods have made significant performance improvement, however, most of them only focus on the single-input single-output (SISO) channel with additive white Gaussian noise (AWGN). Thus, to design a more realistic and practical system for real communication scenarios, it is necessary to consider the MIMO scenario. Therefore, in this paper, we first propose a DL-based JSCC system for image transmission, and extend it into a more practical scenario by taking the MIMO channels into account. Moreover, we consider the precoding design for the physical layer communication in D-JSCC to improve the performance. 
	
	To the best of our knowledge, the data-driven D-JSCC model for image transmission generally achieves different performance among the samples. It is mainly because the complexity, e,g., complex texture, varies among different images. 
	 We notice the images with higher performance are generally with a higher tolerance for the disturbance, i.e., these images are more robust against noise. Intuitively, this phenomenon inspires us to determine the optimal tradeoff between allocated transmission resources and image reconstruction quality for the D-JSCC with MIMO. Hence, if the reconstruction quality can be evaluated in advance, the resource allocation strategy can be determined. To this end, we first develop a lightweight model to predict the performance on each image. It takes the images as input, and outputs the reconstruction quality, e.g., peak signal-to-noise (PSNR).
	 Then, by taking the source image's predicted reconstruction quality into account, we develop an adaptive scheme to adjust the number of feedback bits. In particular, the receiver would determine the compression ratio of the CSI matrix based on the predicted reconstruction quality of the transmitted images, and sends the compressed CSI through the feedback link to the transmitter for precoding. Our simulation results show that the proposed scheme can significantly improve the image transmission performance and effectiveness with much-reduced feedback overhead. 
		
	The rest of this paper is structured as follows. Section \ref{System} introduces the framework of D-JSCC with MIMO. The proposed adaptive CSI feedback scheme based on the predicted performance is presented in Section \ref{propose}. Simulation results are presented in Section \ref{Simulation}. Finally, Section \ref{Conclusion} concludes this paper.



\section{Deep Joint Source Channel Coding with MIMO} \label{System}

In this section, we propose the framework of the practical JSCC system with the MIMO channel.

\subsection{System Model}
The proposed D-JSCC is an end-to-end communication system developed to incorporate the channel coding and source coding. In particular, the encoding, decoding, and transmission procedures are parameterized by the NNs, and the system is optimized in a back-propagation manner with the data-driven method. 

\begin{figure}[t]
	\begin{centering}
		\includegraphics[width=0.46\textwidth]{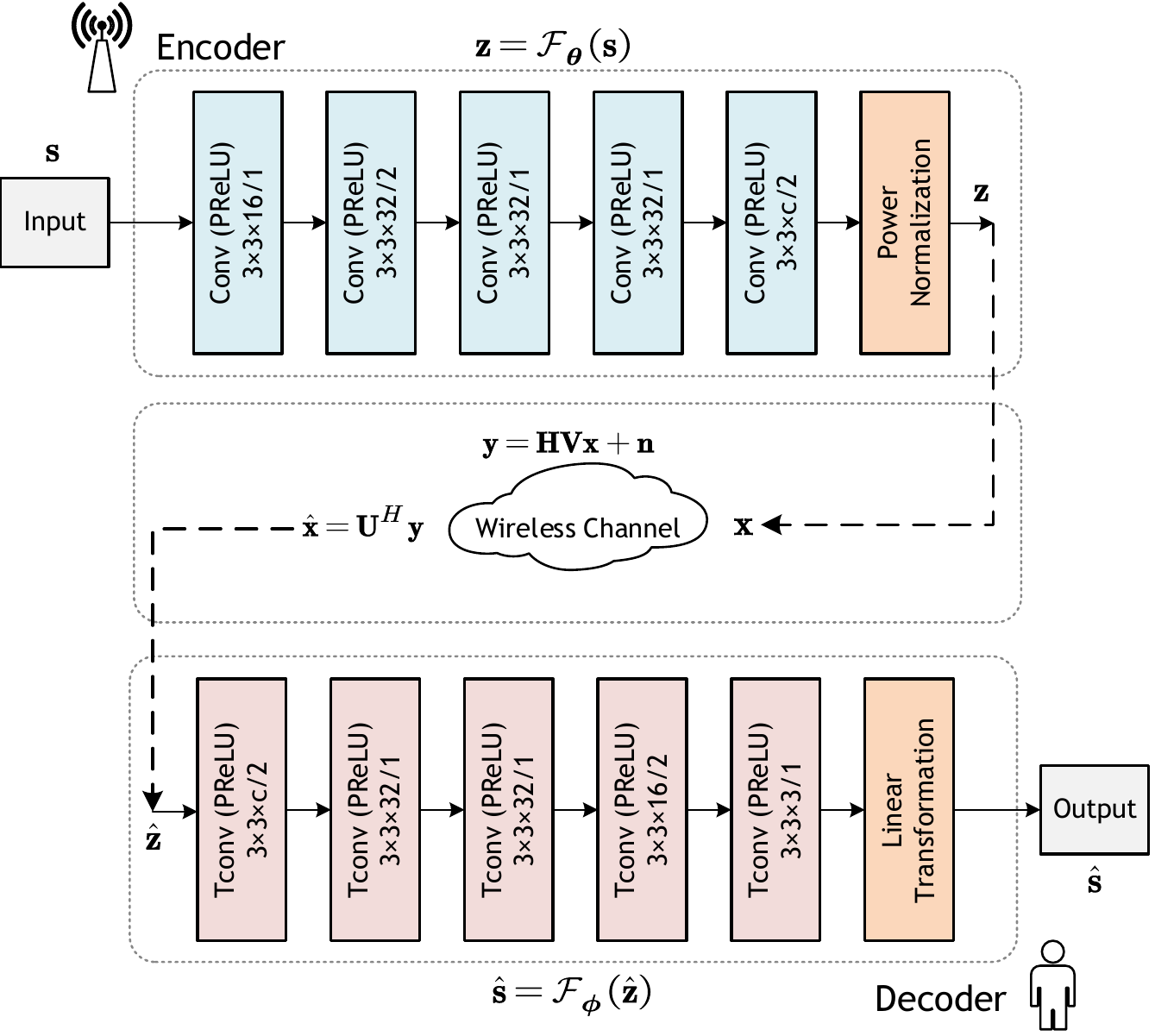}
		\par\end{centering}
	\caption{The architecture of D-JSCC with MIMO channel.}
	\label{DJSCC_Framework}
\end{figure}

\begin{figure*}[t]
	\begin{centering}
		\includegraphics[width=0.7\textwidth]{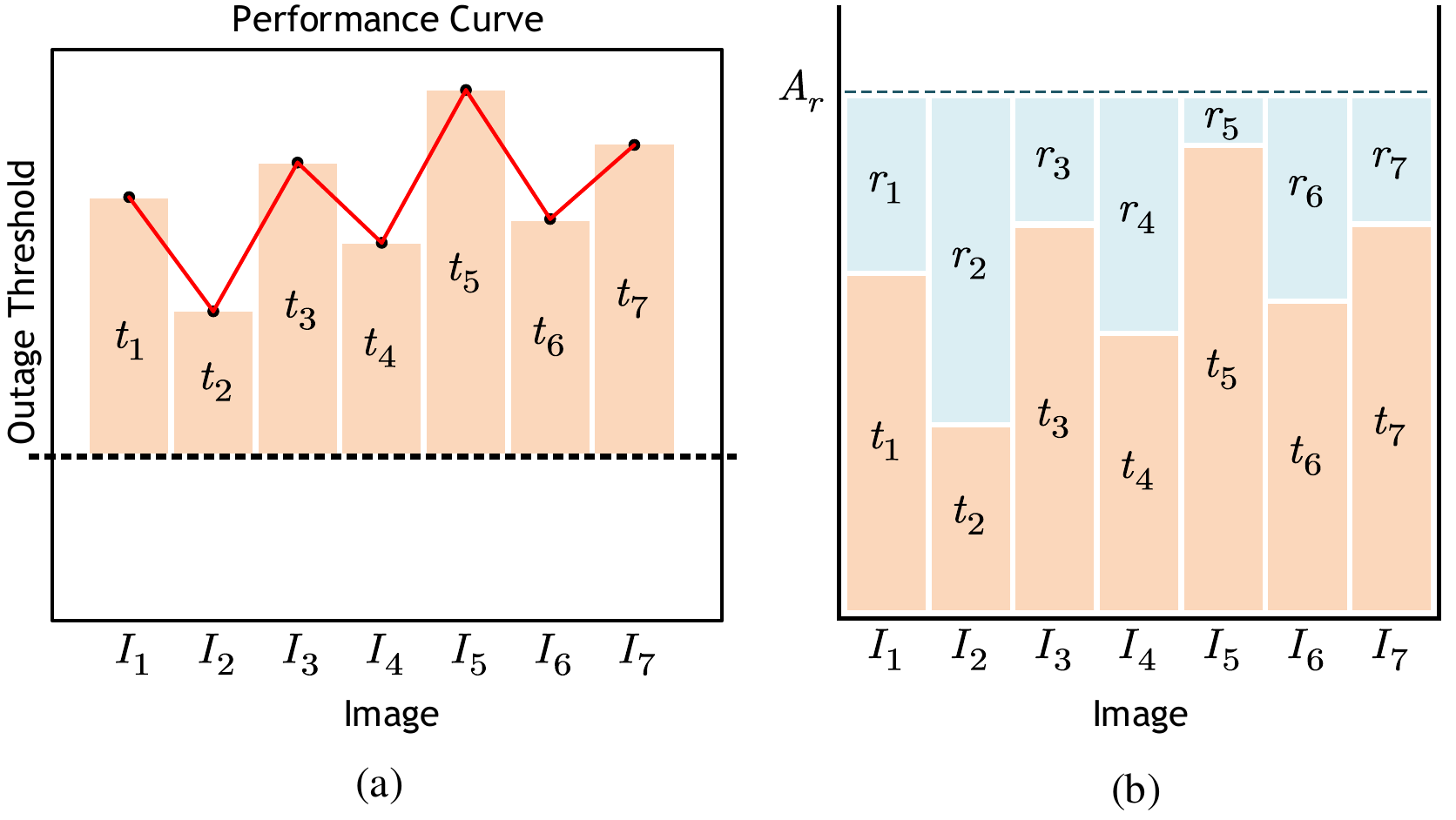}
		\par\end{centering}
	\caption{Visualization of the transmission resources allocation in D-JSCC. (a) The outage threshold and tolerance; (b) Illustration of the water-filling in D-JSCC.}
	\label{ProImage}
\end{figure*}

As shown in Fig. \ref{DJSCC_Framework}, the D-JSCC mainly consists of three parts: encoder, communication channel, and decoder. The encoder and decoder are represented by the DNNs. The input image is represented by a vector, $\mathbf{s}\in \mathbb{R}^{N\times1}$, where $N$ is the length of the vector. Denote the encoding function of the encoder as $\mathcal{F}_{\boldsymbol{\theta}}:\mathbb{R}^{N\times1}\rightarrow \mathbb{C}^{K\times1}$, with $\boldsymbol{\theta}$ denoting its trainable parameters. The encoder encodes $\mathbf{s}$ directly into the complex channel symbol vector, which is given by
\begin{equation}
	\mathbf{z}=\mathcal{F}_{\boldsymbol{\theta}}\left( \mathbf{s}  \right) \in \mathbb{C}^{K\times1},
\end{equation}
where $K$ is the number of transmitted symbols. Then, the symbol vector $\mathbf{z}$ is constrained to satisfy the average power constraint before transmission.

Subsequently, $\mathbf{z}$ is transmitted through the MIMO channel. In particular, we consider a frequency-division duplexing (FDD) system, and the images are transmitted by the base station (BS), which is equipped with $N_t$ transmit antennas. The receiver is equipped with $N_r$ receive antennas. 

Let $\mathbf{V}\in \mathbb{C}^{N_r \times d}$ denote the precoder that BS uses to transmit the signal $\mathbf{s}$, where $d$ denotes the number of data streams. It can be obtained by applying the singular value decomposition (SVD) or zero-forcing (ZF) precoding algorithms with the channel matrix, $\mathbf{H}\in \mathbb{C}^{N_r \times N_t}$, which is obtained via the feedback link from the receiver. Note that the image is encoded by the encoder into a $K$-dimension vector, i.e., the complex channel symbols. We split them into a number of signals, whose dimension all equal to $d$. Take one of the split signal, $\mathbf{x}\in \mathbb{C}^{d \times 1}$, as an example, the received signal can be denoted as
\begin{equation}
	{\mathbf{y}}=\mathbf{H}\mathbf{V}\mathbf{x} + \mathbf{n},
\end{equation}
where $\mathbf{n}\in \mathbb{C}^{N_r \times 1}$ is the AWGN.

At the receiver, we consider linear receive precoding, thus the estimated signal is obtained by 
\begin{equation}
	{\hat{\mathbf{x}}}=\mathbf{U}^{H}\mathbf{y},
\end{equation}
where the receive precoder, $\mathbf{U}$, is obtained by employing the precoding algorithms with $\mathbf{H}$, and $H$ denotes the conjugate transpose. Correspondingly, after receiving these split signals, we obtain the received symbol vector, $\hat{\mathbf{z}}$, which will be further processed by the decoder. The decoder employs the decoding function, $\mathcal{F}_{\boldsymbol{\phi}}:\mathbb{C}^{K\times1}\rightarrow \mathbb{R}^{N\times1}$, to map $\hat{\mathbf{z}}$ into an estimate of the original signal for reconstruction, which is given by
\begin{equation}
	\hat{\mathbf{s}}=\mathcal{F}_{\boldsymbol{\phi}}\left( \hat{\mathbf{z}}  \right) \in \mathbb{R}^{N\times 1},
\end{equation}
where $\boldsymbol{\phi}$ denotes the trainable parameters of the decoder.

\section{Adaptive CSI Feedback for D-JSCC}  \label{propose}
In this section, we design the adaptive CSI feedback scheme based on the predicted reconstruction quality of the image.
\subsection{Reconstruction Performance Prediction}
We have observed that D-JSCC typically achieves different reconstruction qualities on different input images. As for semantic communication, guaranteeing the average reconstruction quality, i.e., average PSNR, is not always suitable, since some of the images would be reconstructed with a rather lower PSNR than the others. In this case, the model performance would not be acceptable for some images. 


Concretely, as shown in Fig. \ref{ProImage} (a), we consider the case that all of the transmitted samples are required to be reconstructed to surpass the given minimum threshold, which can be considered as the outage threshold in D-JSCC. The images with higher reconstruction qualities are generally with a higher tolerance for the disturbance, i.e., these images are more robust. This inspires us that there is a trade-off between the allocated transmission resources and the reconstruction qualities based on the predicted reconstruction quality of the image, and to re-determine the transmission resource allocation for the D-JSCC system. More clearly, as shown in Fig. \ref{ProImage}(a), the tolerance of the image $I_{i}$ is denoted as $t_i$, which can be estimated by the gap between the performance of the image and the given outage threshold. Then, we can allocate the transmission resources, $r_i$, in a similar way to the water-filling method, which is shown in Fig. \ref{ProImage}(b). Specifically, the image with higher tolerance would be allocated with less resources.
	
Therefore, if we can predict the reconstruction quality of the image in advance, it is possible to determine the resource allocation strategy, and improves the efficiency of the model. To achieve this, we firstly develop a performance evaluator. It takes the transmitted images as input and outputs the predicted PSNR value. Compared with the D-JSCC, it is a light model that consists of a few convolutional layers. We model the prediction task as a regression problem, specifically, the PSNR values achieved by the D-JSCC are set as the labels to train the evaluator. 
We exploit the convolutional layer for the architecture of the performance evaluator, which is parameterized by $\mathcal{E}_{\bm{\pi}}$ with $\bm{\pi}$ denoting its trainable parameters. Denote the output predicted PSNR and true achieved PSNR for image $I_i$ as $\hat{\bm{\gamma_i}}$ and $\bm{\gamma_i}$, respectively. The training target for the proposed performance evaluator can be formulated into a regression problem, that is
\begin{equation}\label{L}
	\min_{\bm{\pi}} \  \mathcal{L}=\frac{1}{N}\sum_{i=1}^N (\hat{\bm{\gamma_i}}-\bm{\gamma_i})^{2}=\frac{1}{N}\sum_{i=1}^N \left( \mathcal{E}_{\bm{\pi}}\left(I_i\right)-\bm{\gamma_i}) \right)^{2},
\end{equation}
where $N$ denotes the size of the training batch.


\begin{figure*}[!htbp]
	\begin{centering}
		\includegraphics[width=0.86\textwidth]{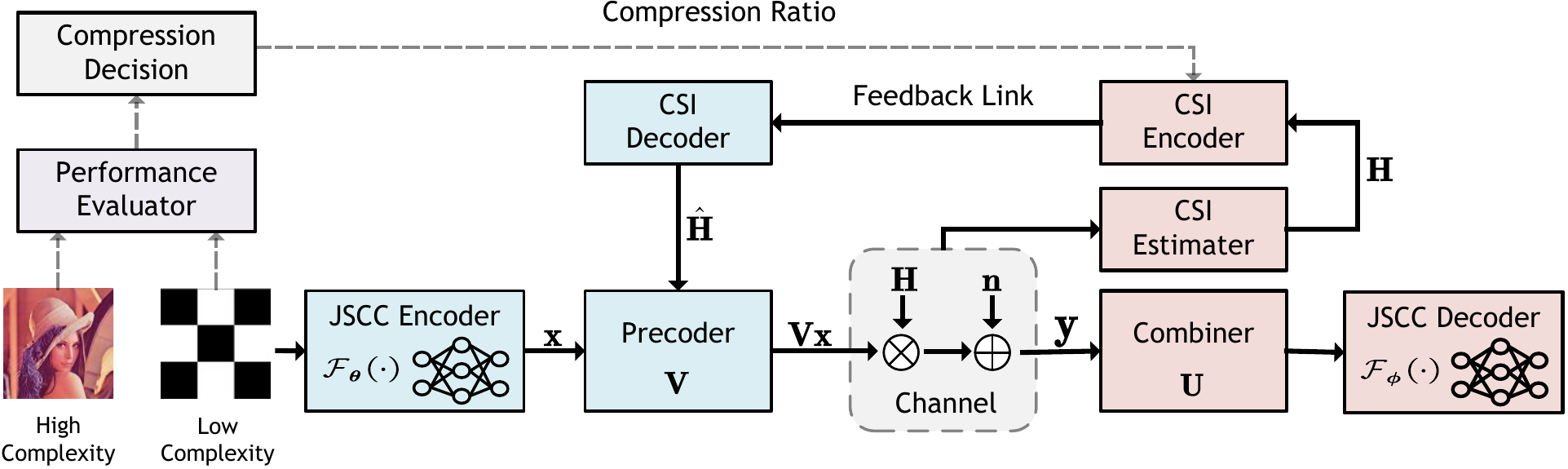}
		\par\end{centering}
	\caption{Framework of the proposed adaptive CSI feedback scheme.}
	\label{CSI_Framework}
\end{figure*}

\subsection{Source-Related Adaptive CSI Feedback}
We have designed the D-JSCC with MIMO in Section \ref{System}, and we aim at adaptively adjusting the overhead of MIMO channel feedback for D-JSCC based on the predicted reconstruction quality. We assume that perfect CSI has been acquired through pilot-based training. As shown in Fig. \ref{CSI_Framework}, the transmitter firstly decides on the compression ratio based on the predicted PSNR value of the image, and sends the compression ratio to the receiver. Then, the estimated real channel, $\mathbf{H}\in \mathbb{C}^{N_r \times N_t}$, will be compressed for feedback based on the compression ratio. Then, the transmitter needs to recover the CSI matrix with the compressed representation to obtain the recovered CSI matrix, $\hat{\mathbf{H}}$. The difference between the recovered CSI matrix, $\hat{\mathbf{H}}$, and real CSI matrix, $\mathbf{H}$ is measured by the normalized mean squared error (NMSE) \cite{CSIPaper1}, which can be computed as 
\begin{equation}
	\mathrm{NMSE}=\mathbb{E}\left\{\frac{\|\mathbf{H}-\hat{\mathbf{H}}\|_2^2}{\|\mathbf{H}\|_2^2}\right\}.
\end{equation}
In particular, the higher compression ratio will lead to the worse reconstruction quality, i.e., a higher NMSE. In traditional MIMO communication systems, the accuracy of the CSI feedback has an impact on the bit error rate (BER). Similarly, the difference between $\mathbf{H}$ and $\hat{\mathbf{H}}$ can decrease the performance of D-JSCC. Although increasing the compression ratio helps to reduce the difference, it will induce high feedback overhead. Thus, there is a trade-off between the feedback overhead and system performance. We aim to reduce the feedback overhead by adjusting the compression rate of the CSI matrix for different images. Specifically, as shown in Fig. \ref{CSI_Framework}, the transmitter decides on the compression ratio based on the predicted PSNR values of the given images. If the images have low predicted PSNR values, the corresponding CSI matrix will be allocated with more feedback bits. On the other hand, the corresponding CSI matrix will be allocated with fewer feedback bits for the images with high predicted PSNR values.

\subsection{Rethinking the Outage in D-JSCC}
In addition to the BER, the outage probability $P_{out}$ is another performance measure of communication systems over fading channels. It is defined as the probability that the (instantaneous) combined signal-to-noise ratio (SNR) $\gamma _{t}$ falls below a certain given threshold $\gamma_{th}$, i.e., 
\begin{equation}
	P_{ {out }}=P\left[0 \leq \gamma_t \leq \gamma_{\text {th}}\right]=\int_0^{\gamma_{\text {th }}} p_{\gamma_t}\left(\gamma_t\right) d \gamma_t,
\end{equation}
where $p_{\gamma_t}\left(\gamma_t\right) $ denotes the probability density function of $\gamma_t $. In other words, $P_{out}$ is the cumulative distribution function of $\gamma_{t}$ evaluated at $\gamma_{th}$.

 
 
However, the outage in the traditional communication system is not applicable in D-JSCC since D-JSCC only considers the reconstruction quality rather than the BER. Therefore, we aim to redefine the outage. In particular, we define the outage in D-JSCC as the reconstruction quality falls below a certain given threshold, and the images are required to be reconstructed with a quality that surpasses the given outage threshold in practical D-JSCC systems. Correspondingly, the outage probability in D-JSCC can be evaluated by the ratio of images that are reconstructed with quality less than the given outage threshold. Moreover, the reconstruction quality is dependent on both the SNR and the complexity of the image in D-JSCC, which is quite different from that in traditional communication systems.

In fact, as for human perception, when the reconstruction quality reaches a certain threshold, the difference between the transmitted image and the reconstructed one is imperceptible. Moreover, there are also scenarios where the BS reduces the data quality, e.g., the quality of image and video, to a low level to save the resources to serve more users. It is equivalent to reducing the threshold of the system. Therefore, it is important and practical to define the outage for D-JSCC to ensure its reliability and effectiveness in real communication systems.

\subsection{Training Method}

To jointly learn the encoder and decoder with back-propagation, we employ the mean square-error (MSE) loss, which is given by
\begin{equation}
	\mathcal{L}\left(\mathbf{s},\hat{\mathbf{s}}\right)=\frac{1}{N} \sum_{i=1}^N\left(s_i-\hat{s}_i\right)^2,
\end{equation}
where $s_i$ and $\hat{s_i}$ denote the corresponding $i$-th element of the $\mathbf{s}$ and $\hat{\mathbf{s}}$, respectively. 

\section{Simulation Results} \label{Simulation}
In the simulations, we consider a transmitter equipped with $N_t=16$ transmit antennas, and a receiver equipped with $N_r=16$ receive antennas. The number of streams, $d$, is $2$. We implement the proposed D-JSCC and the DL-based CSI feedback scheme with the deep learning platform “Pytorch”. The “Adam” optimizer is employed as the optimizer, with a batch size of $128$. Moreover, the initial learning rate is $0.001$ and will be reduced with the increase of the number of epochs. For simulation, we use the CIFAR10 dataset which consists of $50000$ color images of size $32\times32\times3$ in the training dataset and $10000$ images in the test dataset.

We employ the popular narrowband millimeter wave (mmWave) clustered channel \cite{mmWave}, where the numbers of clusters and propagating rays are $N_{cl}$ and $N_{ray}$, respectively. According to \cite{mmWave}, the channel matrix is given by  
\begin{equation}
	\mathbf{H}=\sqrt{\frac{N_t N_r}{N_{c l} N_{\text {ray }}}} \sum_{i=1}^{N_{c l}} \sum_{l=1}^{N_{r a y}} \alpha_{i l} \mathbf{a}_r\left(\phi_{i l}^r\right) \mathbf{a}_t^H\left(\phi_{i l}^t\right),
\end{equation}
where $ \alpha_{i l}$ denotes the complex gain of the $l$-th ray in the $i$-th cluster. Moreover, $\phi_{i l}^r$ and $\phi_{i l}^t$ are respectively the azimuth angles at the receiver and transmitter for the $l$-th ray in the $i$-th cluster. The $\mathbf{a}_r\left(\phi_{i l}^r\right) $ and $\mathbf{a}_t\left(\phi_{i l}^t\right)$ denote the receive and transmit array response vectors, respectively. Consider a uniform linear array, the response vector is given by 
\begin{equation}
	\mathbf{a}(\phi)=\frac{1}{\sqrt{N}}\left[1, e^{-j 2 \pi \frac{d}{\lambda} \sin (\phi)}, \ldots, e^{-j 2 \pi \frac{d}{\lambda}(N-1) \sin (\phi)}\right]^T,
\end{equation}
where $N$ and $\phi$ are the numbers of antenna elements and azimuth angle, respectively. Moreover, $d$ and $\lambda$ represent the adjacent distance antennas and carrier wavelength, respectively.  In the simulation, we choose $N_{cl}=2$ clusters and $N_{ray}=4$ rays. To quantize the CSI matrix, we employ the non-uniform quantization method, Lloyd-Max algorithm, which quantizes the parameter of the CSI matrix with a generated table. In each transmission, we randomly generate a channel matrix for each image, where we consider a block-fading scenario. 

To evaluate the performance of the D-JSCC, the PSNR is adopted. 
 It measures the ratio between the maximum possible power and the noise, which can be calculated by
\begin{equation}
	\textrm{PSNR}=10 \log_{10}{\frac{\textrm{MAX}^{2}}{\textrm{MSE}}}(\textrm{dB}),
\end{equation}
where $\textrm{MSE}=d(\mathbf{s},\hat{\mathbf{s}})$ represents the mean quare-error (MSE) between the source image, $\mathbf{s}$, and the reconstructed image, $\hat{\mathbf{s}}$. Moreover, $\textrm{MAX}$ is the maximum possible value of the pixels, e.g., $\textrm{MAX}$ equals $255$ for the images of RGB format.

\begin{figure}[t]
	\begin{centering}
		\includegraphics[width=0.40\textwidth]{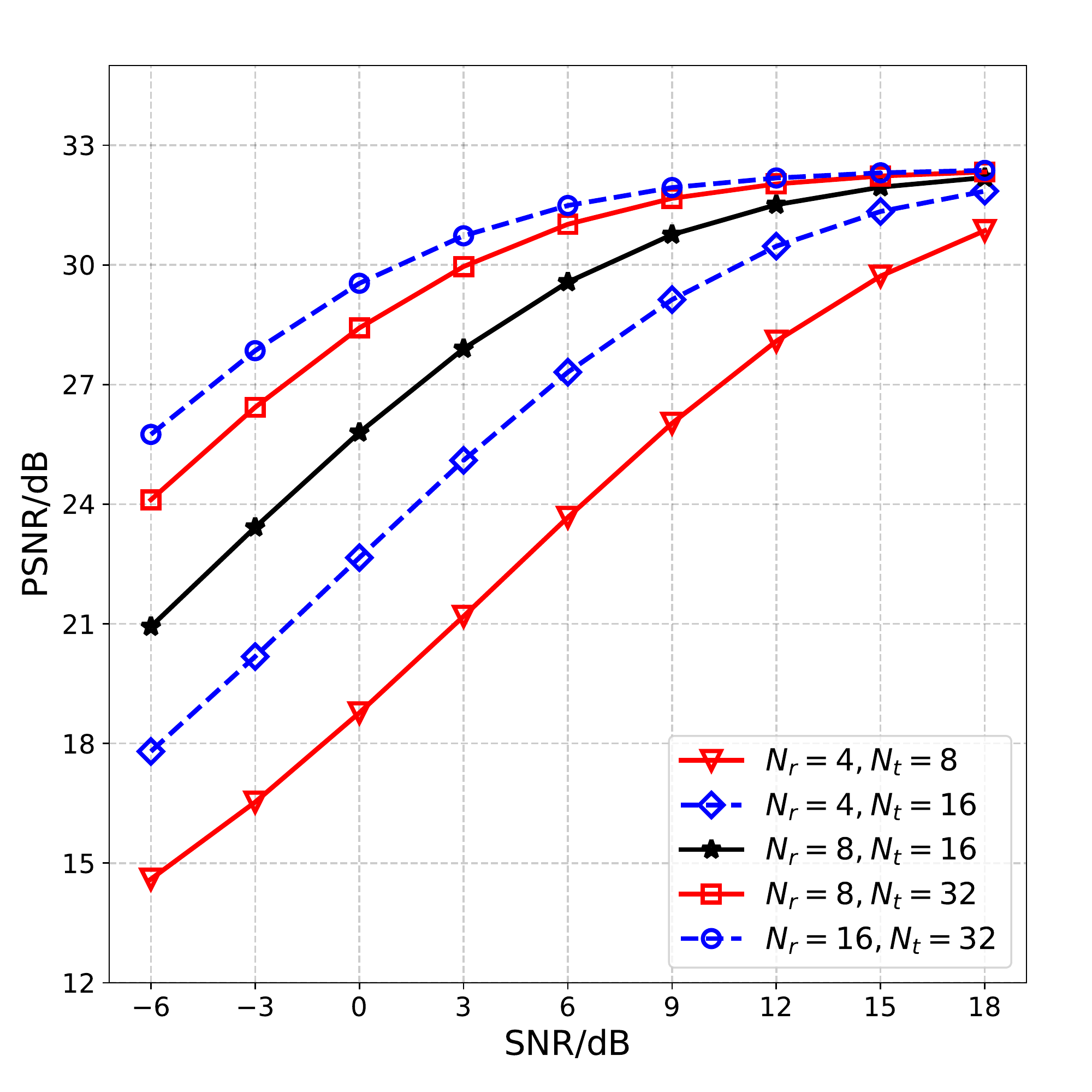}
		\par\end{centering}
	\caption{The performance of proposed model with different numbers of antennas versus SNR.}
	\label{MIMO_Sim}
\end{figure}

Fig. \ref{MIMO_Sim} presents the performance of the investigated schemes equipped with different numbers of antennas versus the SNR, and we assume that the transmitter obtains the perfect CSI. We train the proposed D-JSCC model with SNR = $6$ dB and test it in SNR from $-6$ dB to $18$ dB. It is readily seen that the PSNR achieved by D-JSCC increases with SNR. The system equipped with more antennas generally outperforms the system with fewer antennas. It is mainly because the diversity gain increases with the number of antennas. It demonstrates that the MIMO system is still effective in improving the performance of DL-based D-JSCC.

\begin{figure}[t]
	\begin{centering}
		\includegraphics[width=0.40\textwidth]{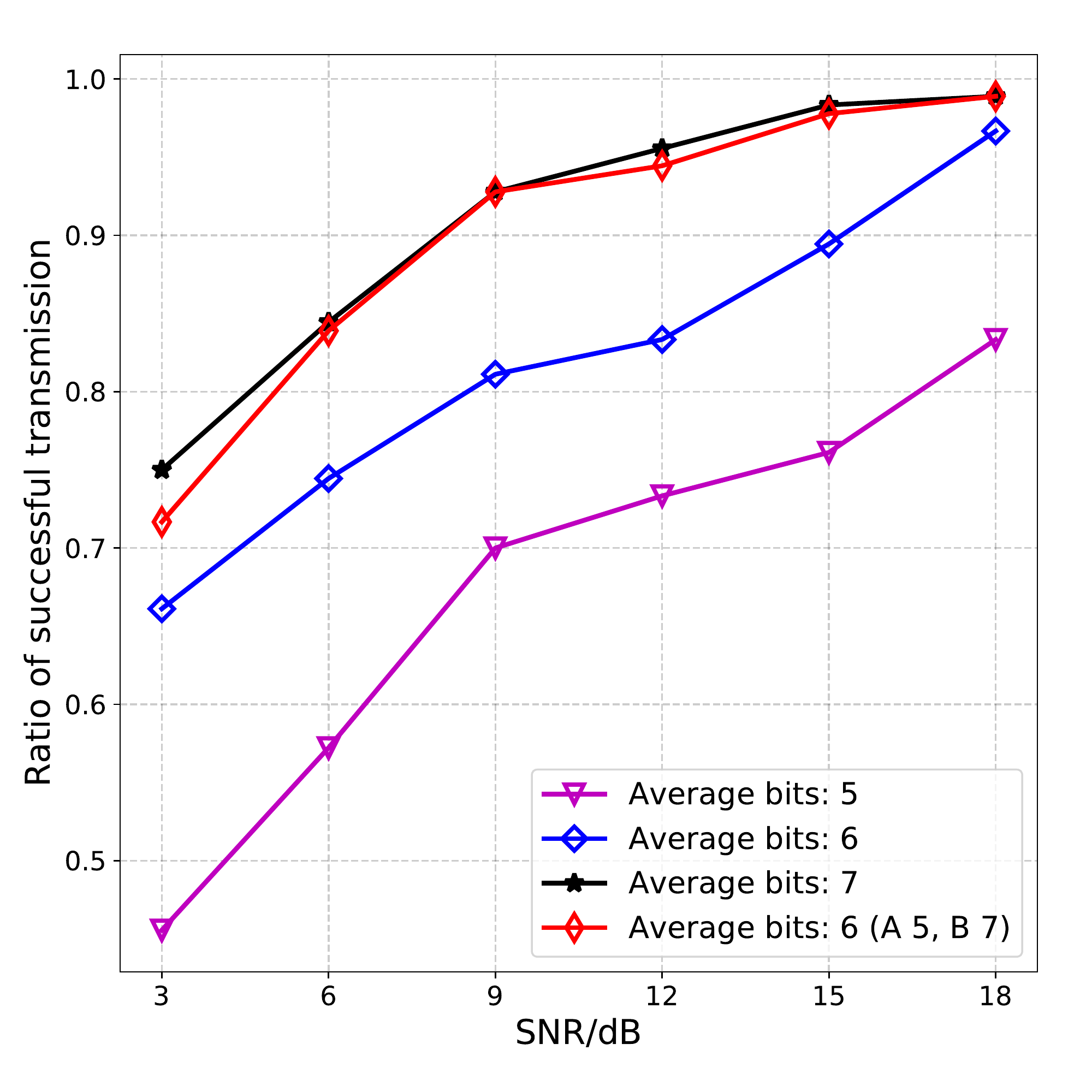}
		\par\end{centering}
	\caption{The performance of the proposed adaptive scheme with different numbers of feedback bits.}
	\label{CSI_Sim}
\end{figure}

Fig. \ref{CSI_Sim} presents the ratio of successful transmission of the proposed adaptive scheme with different numbers of feedback bits. In particular, the ratio is obtained by dividing the number of images that have higher reconstruction quality (PSNR) than the threshold by the total number of testing images. Moreover, the number of feedback bits refers to the number of the average quantization bits for each element in the channel matrix. As for the red curve, we divide the transmitted images into two groups, A and B, according to their predicted PSNR values. The numbers of images in the group with higher predicted PSNR values, A, and the group with lower predicted PSNR values, B,  are the same. To make the number of average feedback bits for A and B equal to $6$, we set the number of feedback bits for A as $5$, and that for B as $7$, thus the average number is $6$.  As for other curves, the numbers of feedback bits for all the images are both fixed at the same value, e.g., $5$ for the blue curve. From the figure, we can see that allocating the same feedback bits for each image is not optimal, since the reconstruction qualities of the images are different. In comparison, our proposed adaptive CSI feedback scheme significantly outperforms the average strategy by adaptively adjusting the number of feedback bits. 

\begin{figure}[t]
	\begin{centering}
		\includegraphics[width=0.40\textwidth]{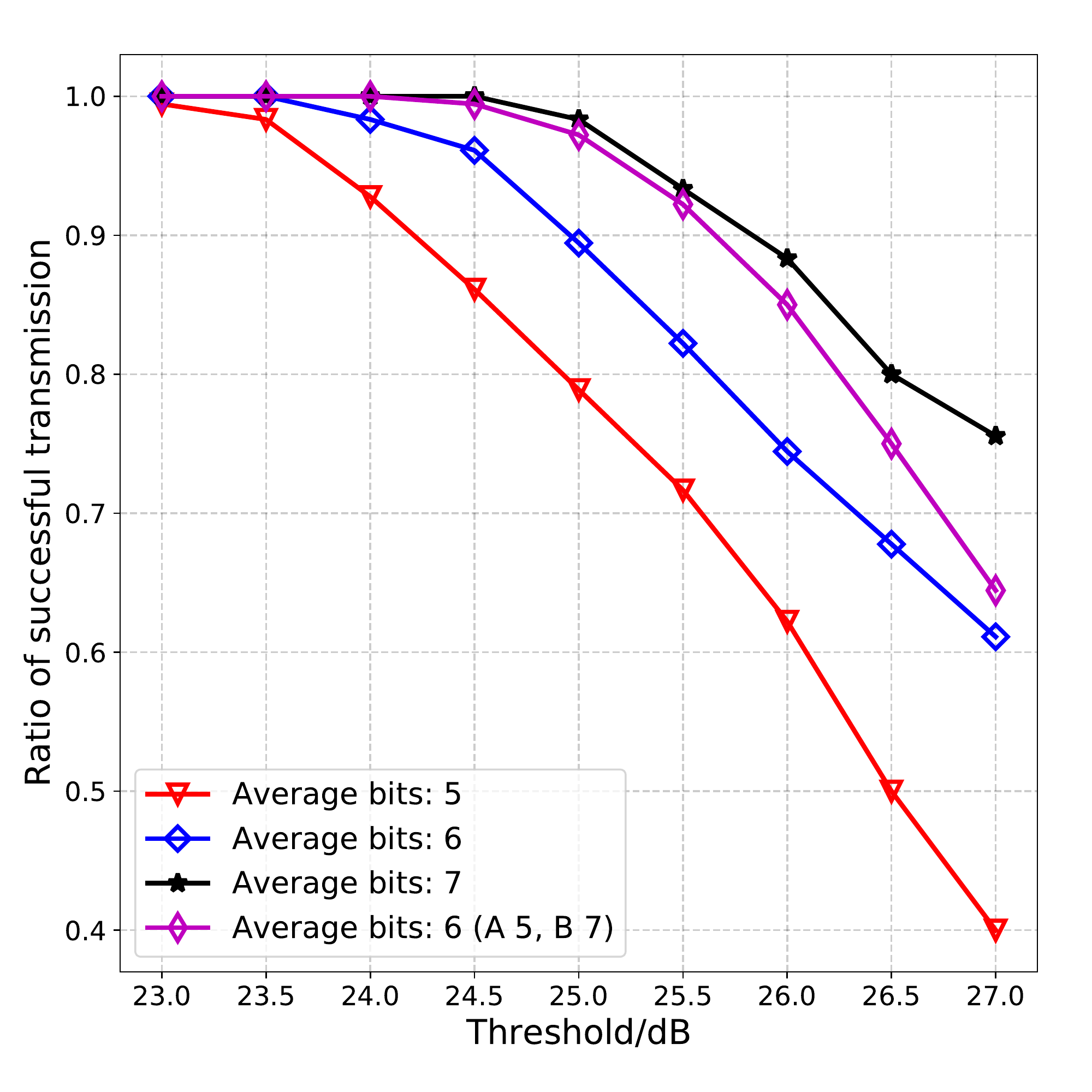}
		\par\end{centering}
	\caption{ The performance of the proposed adaptive scheme versus the threshold.}
	\label{CSI_Sim_same}
\end{figure}
Fig. \ref{CSI_Sim_same} shows the ratio of successful transmission of the proposed adaptive scheme versus the outage threshold. In this figure, all the schemes are required to meet the given outage threshold. From the figure, we can see that the performance of these schemes decrease with the threshold. Moreover, our proposed scheme outperforms the scheme with the same feedback bits, as shown by the blue line. It demonstrates the superiority of the proposed adaptive schemes in different threshold requirements.  

\begin{figure}[t]
	\begin{centering}
		\includegraphics[width=0.40\textwidth]{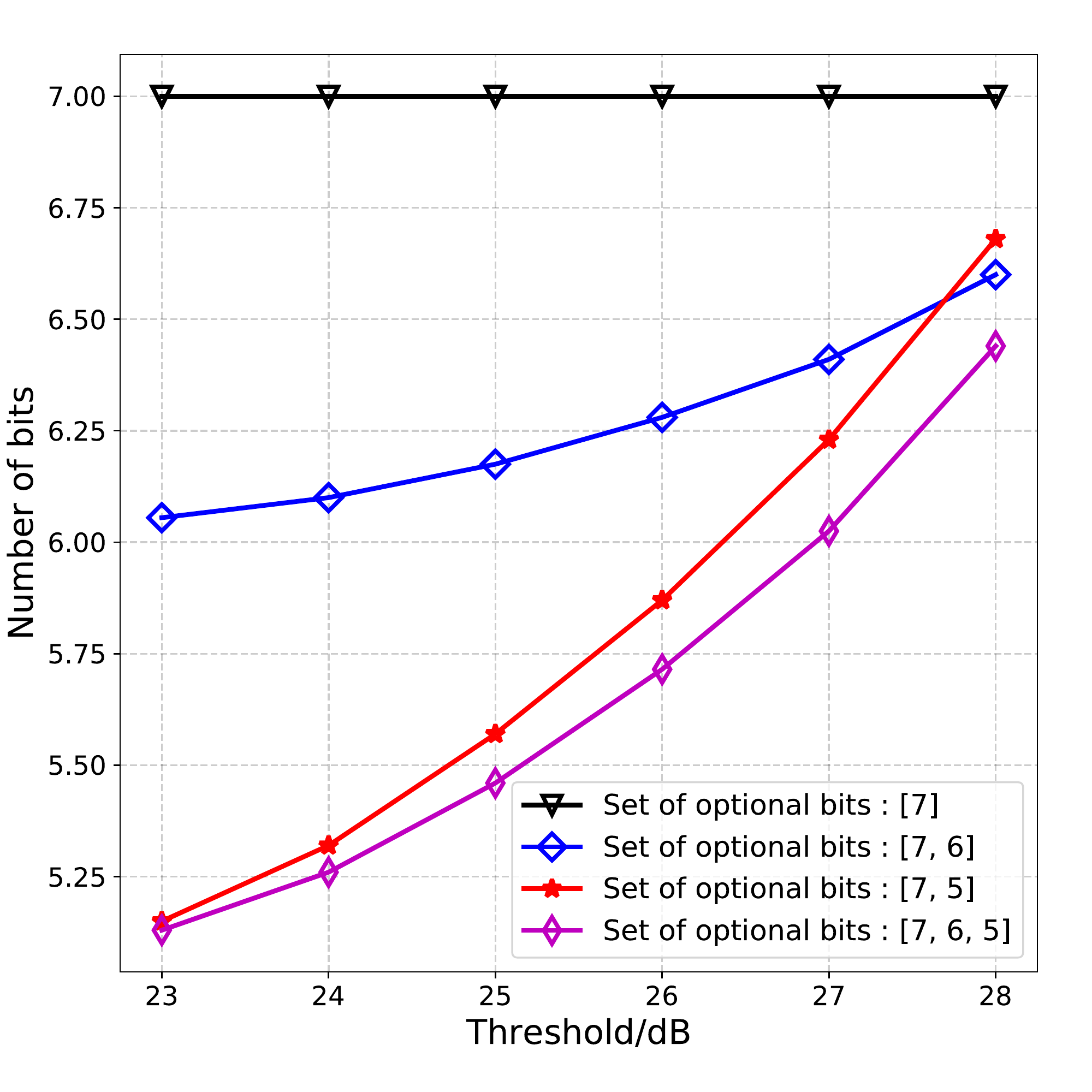}
		\par\end{centering}
	\caption{ The number of required bits versus the threshold.}
	\label{CSI_Numbits}
\end{figure}

In Fig. \ref{CSI_Numbits}, we compare the number of required feedback bits when keeping the same ratio of successful transmission. Specifically, all the schemes are required to achieve the same performance as that $7$ bits feedback is used for all the test images. We counted the number of required bits when the set of optional bits are given, i.e., all the images need to be transmitted with the number of feedback bits selected from the given set. For example, as for $[7,6]$, the images should be allocated with $7$ or $6$ feedback bits for transmission. From the figure, it is readily seen that our proposed adaptive scheme can reduce the feedback overhead while keeping the same ratio of successful transmission. Moreover, with more optional numbers, the feedback overhead can be further reduced.

\section{Conclusion} \label{Conclusion}
In this paper, we have proposed a D-JSCC communication system for image transmission over MIMO channels. Then, to improve the performance, we have designed the precoding for the D-JSCC. The encoder and decoder are jointly trained in an end-to-end manner. Moreover, we have redefined the outage in the D-JSCC and propose an adaptive CSI feedback scheme. It is able to adjust the compression rate based on the predicted performance of the image. Simulation results show that the proposed scheme can significantly improve the image transmission performance and effectiveness with significantly reduced feedback overhead.

\bibliographystyle{IEEEtran}
\bibliography{IEEEabrv,Reference}

\end{document}